\documentclass[twocolumn,showpacs,preprintnumbers,amsmath,amssymb]{revtex4}
\usepackage{mathrsfs}%====================================
\usepackage{amssymb}
\usepackage{graphicx}% Include figure files
\usepackage{dcolumn}% Align table columns on decimal point
\usepackage{bm}% bold math
\usepackage{textcomp}
\usepackage{amsmath}
\usepackage[titletoc]{appendix}

\newcommand\ket[1]{\ensuremath{|#1\rangle}}

\newcounter{RomanNumber}

\begin{document}
\title{Sending or not sending: Twin-field quantum key distribution with large misalignment error
}
\author{Xiang-Bin Wang,$ ^{1,2,4\footnote{Email
Address: xbwang@mail.tsinghua.edu.cn}} $
}

%\email[]{xbwang@mail.tsinghua.edu.cn}

\author{Zong-Wen Yu$ ^{3}$, and Xiao-Long Hu$ ^{1}$}

\affiliation{ \centerline{$^{1}$State Key Laboratory of Low
Dimensional Quantum Physics, Department of Physics,} \centerline{Tsinghua University, Beijing 100084,
People¡¯s Republic of China}
\centerline{$^{2}$ Synergetic Innovation Center of Quantum Information and Quantum Physics, University of Science and Technology of China}
\centerline{  Hefei, Anhui 230026, China
 }
\centerline{$^{3}$Data Communication Science and Technology Research Institute, Beijing 100191, China}
\centerline{$^{4}$ Jinan Institute of Quantum technology, SAICT, Jinan 250101,
People¡¯s Republic of China}}
%%%%%%%%%%%%%%%%%%%%%%%%%%%%%%%%%%%%%%%%%%%%%%%%%%%%%%%%%%%%%%%%%%%
%%%%%%%%%%%%%%%%%%%%%%%%%%%%%%%%%%%%%%%%%%%%%%%%%%%%%%%%%%%%%%%%%%%
%%%%%%%%%%%%%%%%%%%%%%%%% Abstract %%%%%%%%%%%%%%%%%%%%%%%%%%%%%%%%
\begin{abstract}
Based on the novel idea of twin-field quantum key distribution (TF-QKD, by  M. Lucamarini ,Z.L. Yuan, J.F. Dynes, \& A.J. Shields, Nature 557, pages 400-403 (2018)), we present a protocol named as ``sending or not sending TF-QKD" protocol which can tolerate large misalignment error.
  The revolutionary theoretical breakthrough in quantum communication, TF-QKD, changes the channel-loss dependence of the key rate from linear to square root. However, it demands the challenging technology of long distance single-photon interference, and also, as stated in the original paper, the security proof was not finalized there due to the possible effects of the afterwards announced phase information. Here we  show by a concrete Eavesdropping scheme
  that the afterwards phase announcement do have important effects and the traditional formulas of decoy-state method does not apply to the original protocol. We then present our ``sending or not sending" protocol.
Our protocol does not take post selection for the bits in $Z$ basis (signal pulses) and hence the traditional decoy-state method directly apply therefore  automatically resolves the issue of security proof. Most importantly, our protocol presents a negligibly small error rate in $Z$-basis because it does not request any single-photon interference in this basis. This makes our protocol greatly improve the tolerable  threshold of misalignment error  in single-photon interference from the original a few percent to more than $45\%$.    As shown numerically, our protocol exceeds a secure distance of 700 km,  600 km, 500 km, or 300 km even though the single-photon interference misalignment error rate is as large as $15\%$, $25\%$, $35\%$, or $45\%$.    \end{abstract}

%%%%%%%%%%%%%%%%%%%%%%%%%%%%%%%%%%%%%%%%%%%%%%%%%%%%%%%%%%%%%%%%%%%
%%%%%%%%%%%%%%%%%%%%%%%%%%%%%%%%%%%%%%%%%%%%%%%%%%%%%%%%%%%%%%%%%%%
%%%%%%%%%%%%%%%%%%%%%%%%%%%%%%%%%%%%%%%%%%%%%%%%%%%%%%%%%%%%%%%%%%%

\maketitle

%%%%%%%%%%%%%%%%%%%%%%%%%%%%%%%%%%%%%%%%%%%%%%%%%%%%%%%%%%%%%%%%%%%
%%%%%%%%%%%%%%%%%%%%%%%%%%%%%%%%%%%%%%%%%%%%%%%%%%%%%%%%%%%%%%%%%%%
%%%%%%%%%%%%%%%%%%%%%%%%%%%%%%%%%%%%%%%%%%%%%%%%%%%%%%%%%%%%%%%%%%%
%%%%%%%% Introducation & Motivation %%%%%%%%%%%%%%%%%%%%%%%%%%%%%%%

\section{ Introduction}
Quantum key distribution (QKD) \cite{BB84,GRTZ02} can in principle present secure private communications with its security guaranteed by principles of quantum physics. With the development \cite{curty1,ind3,H03,wang05,LMC05,tittel,liu1,a11,t16,u16,g17,nature,gllp} in both theory and experiment, it is more and more hoped to be extensively applied in practice, though there are barriers for so. Among all barriers, channel loss of long distance QKD is the major one \cite{a11,u16}.
\\
Very recently,  a revolutionary theoretical progress was made by Lucamarini et al. They proposed the novel idea of twin-field quantum key distribution (TF-QKD) \cite{nature} which has historically changed the relationship between key rate and the channel loss from linearly dependent to square root dependent. Consequently, the TF-QKD  makes a great breakthrough of a secure distance longer than 500 km.

 In the twin-field quantum key distribution (TF-QKD)\cite{nature}, Alice and Bob send fields to the un-trusted third party Charlie.   In an virtual ideal protocol, Alice and Bob initially share single-photon entangled states of $|\Phi^0\rangle=\frac{1}{\sqrt 2}(|01\rangle+|10\rangle)$. They each will take a phase shift of either 0 or $\pi$ to each one's local field and they will send their fields to Charlie. After a collective measurement, Charlie will see whether the bipartite is $|\Phi^0\rangle$ or $|\Phi^1\rangle=\frac{1}{\sqrt 2}(|01\rangle-|10\rangle)$. Except Alice and Bob, no one knows which value, 0 or $\pi$ was selected by Alice or Bob in doing their phase shift, although it is known to everyone whether Alice and Bob has used the same phase shift or different phase shift.
So, they can use the information whether Alice has taken a phase shift 0 or $\pi$ for their secret key.

However, in practice, we do not have such an initially shared states. The TF-QKD proposed to use weak coherent states at each side.
As was stated in the original article \cite{nature}, the security is not finally completed because the possible effects of afterwards announcement of the phase information are not taken into consideration. As shown by a concrete Eavesdropping scheme in the supplement, we find that the phase information announced afterwards makes the traditional formulas of the decoy state method \cite{H03,wang05,LMC05} do not apply to the original protocol \cite{nature}. In fact£¬ given the scheme in the appendix, Eve can have full information to the key bits while traditional decoy-state method can give a key rate of $50\%$. Our Eavesdropping scheme shows that the fraction of single-photon bits among all raw bits must be not less than $50\%$, otherwise Eve may have full information to all bits without causing any disturbance.  Although one may naturally turn to the key rate  formulas for non-random-phase coherent states to resolve the issue, however,  TF-QKD relied on the challenging technology of long distance single-photon interference, which may produce large misalignment error.  Here we  construct a `` sending or not sending "  TF-QKD  protocol where there is no phase-slice dependent post selection for signal bits. Not only this itself increases the amount of key bits, but also, this makes  the traditional calculation formulas for the decoy state method directly apply, the security proof is automatically completed and the less efficient key rate formula for non-phase-random coherent states is not necessary. Most importantly, our protocol can tolerate large misalignment error rate due to the long distance single-photon interference.
\section{Sending or not-sending (SNS) protocol}
\noindent {\bf Step 0}. At any time window $i$, as requested by the TF-QKD, {\em they} (Alice and Bob) take random phase shifts $\delta_{Ai},\delta_{Bi}$ to their coherent states accompanied by the reference light which is sent to Charlie. Charlie is also supposed to do appropriate phase compensation, but he is possibly dishonest.
 \\
{\bf Step 1}. At any time window $i$, Alice (Bob) independently determines whether it is a decoy window or a signal window.
If it is a decoy window, she (he) sends out to Charlie a decoy pulse  in coherent state $|\sqrt\mu e^{i\delta_{Ai}+i\gamma_{Ai}}\rangle$
($|\sqrt\mu e^{i\delta_{Bi}+i\gamma_{Bi}}\rangle$ ); and $\mu$ can randomly change among a few different values at different decoy-windows.
If it is a signal window, she (he) decides to send out to Charlie a signal pulse $|\sqrt{\mu'} e^{i\delta_{Ai}+i\gamma_{Ai}}\rangle$ ($|\sqrt{\mu'} e^{i\delta_{Bi}+i\gamma_{Bi}}\rangle$ )   by probability $\epsilon$  and she (he) decides not to send out anything by probability $1-\epsilon$.
Given whatever window she (he) commits, and whatever decision she (he) makes,   the random phase-shift values of $\delta_{Ai}$ ($\delta_{Bi}$) are taken privately by Alice and Bob, respectively.
Given whatever window she (he) commits, and whatever decision she (he) makes, the global phases $\gamma_{Ai}$  ($\gamma_{Bi}$) are always announced, e.g., by a strong reference light.
\\
Note: This sending by small probability $\epsilon$ or not sending by probability $1-\epsilon$ is the heart of our protocol.
\\
Note: A coherent state of intensity $x$ and global phase $\gamma$ is a linear superposition of photon number states $\{|k\rangle\}$ of
$|\sqrt x e^{i\gamma}\rangle = \sum_{k=0}^\infty \frac{e^{-x/2}(\sqrt x e^{i\gamma})^k}{\sqrt{k!}}|k\rangle$. In a signal window, if Alice or Bob decides to send, she (he) shall always send a coherent state of intensity $\mu'$. For example, at a certain time  when they both determined signal windows, if Alice decides to send while Bob decides not to send, the two-mode state from this time window is $|\sqrt{\mu'} e^{i\delta_A+i\gamma_A}\rangle \otimes |0\rangle$; if both of them decide to send, the two-mode state is $|\sqrt{\mu'} e^{i\delta_A+i\gamma_A}\rangle\otimes
|\sqrt{\mu'} e^{i\delta_B+i\gamma_B}\rangle$; if both of them decide not to send, the state at that time window is $|00\rangle$. Here $\gamma_A,\gamma_B$ are global phases of the coherent states. They are known to Eve because Alice and Bob also send strong reference pulses accompany the weak coherent light. States from a decoy window can have different intensities. If at a certain time both of them have chosen decoy window and both of them have happened to choose the same intensity $\mu$, the two-mode coherent state from this time window is
$|\sqrt{\mu} e^{i\delta_A+i\gamma_A}\rangle\otimes
|\sqrt{\mu} e^{i\delta_B+i\gamma_B}\rangle$. In the protocol, Charlie is supposed to do phase compensation, trying to remove the global phases. If Charlie does this perfectly, the states from each side after the compensation have the same global phases. For example,   state $|\sqrt{\mu} e^{i\delta_A+i\gamma_A}\rangle\otimes
|\sqrt{\mu} e^{i\delta_B+i\gamma_B}\rangle$ will be changed into $|\sqrt{\mu} e^{i\delta_A+i\gamma_A}\rangle\otimes
|\sqrt{\mu} e^{i\delta_B+i\gamma_B}\rangle$ after a perfect phase compensation by Charlie.
  \\
{\bf Step 2}.  Charlie is supposed to measure all twin-fields with a beam-splitter after taking phase compensation  and announce the measurement outcome.
\\
Note:   We define an {\em effective event} by the following criterion:
(i) If Charlie announces only one detector counting corresponding to a time window $i$ when both of them have determined a signal window, it is an effective event; (ii) if Charlie announces only one detector counting corresponding to a time window $i$ when both of them have determined a decoy window, used the same intensity of coherent states, and in that time window, the pre-chosen values $\delta_A,\delta_B$ satisfy
\begin{equation}\label{slick}
1-|\cos (\delta_A-\delta_B)|\le |\lambda|.
\end{equation}
Here the value $\lambda$ is determined by the size of phase slice\cite{nature} chosen by Alice and Bob.
Whenever an effective event happens, a bit in the corresponding basis is recorded. \\
{\bf Step 3}.
{\em They} announce each one's  decoy windows and  signal windows. {\em They} also announce details for intensities of pulses sent from decoy windows and values $\delta_A,\delta_B$ they each have used.
\\
Note: We define a $Z$-window as a time window when both Alice and Bob have determined a signal window. We name states from such $Z$-windows  as states in  $Z$-basis, or simply $Z$-pairs, $Z$-states.  Effective events happen in $Z$-basis are named as $Z$-bits. Given that $\delta_A$ value ($\delta_B$ value) is randomized, whenever Alice or Bob sends a coherent state of intensity $\mu'$, it can be equivalently regarded as a density matrix of $\int_0^{2\pi} |\sqrt {\mu'} e^{i\delta_A+i\gamma_A}\rangle\langle \sqrt {\mu'} e^{i\delta_A+i\gamma_A }|d\delta_A=\sum_{k=0}^\infty \frac{e^{-\mu'}{\mu'}^k}{k!}|k\rangle\langle k|$, which is a classical mixture of different photon number states only. Hence we can define $Z_1$-windows as a subset of  $Z$-windows when only one party of Alice and Bob decides to send and she (he) actually sends a single-photon state. In a $Z_1$-window, the two-mode single photon state sent out is either $|z_0\rangle=|01\rangle$ or $|z_1\rangle =|10\rangle$. We shall call them as  $Z_1$ states or $Z_1$ pairs. Also, effective events caused in  $Z_1$-windows are named as $Z_1$-bits.  Furthermore, we define an $X$-window as a time window when  $1$) both of them have chosen the decoy window, $2$) both of them have chosen the same intensity for the coherent state to send,  and $3$) the random phase $\delta_A$, $\delta_B$ chosen for the window satisfy
Eq.(\ref{slick}). We name the two-mode states from $X$-windows as states in $X$-basis,  or simply $X$-pairs or  $X$-states, and  an  $X$-bit is a bit caused by $X$ pair. Also, as shown later, states of $X$ pairs can be regarded as a probabilistic mixture of different photon-number states, with the two-mode single-photon ingredient $|\psi_1\rangle\langle \psi_1|$, and $|\psi_1\rangle = \frac{1}{\sqrt 2}(e^{i(\delta_B+\gamma_B)}|01\rangle + e^{i(\delta_A+\gamma_A)}|10\rangle)$. Therefore we can define an $\mathcal X_1$-window as an $X$-window when {\em they} send a (two-mode) single-photon state. We also name those states from $\mathcal X_1$-windows  as $\mathcal X_1$-pairs or $\mathcal X_1$-states, and the bits caused $\mathcal X_1$ pairs  as $\mathcal X_1$-bits. {\em They} do not know which time windows are $Z_1$-windows and $\mathcal X_1$-windows, neither  do {\em they} know which bits are $Z_1$-bits and $\mathcal X_1$-bits, though {\em they} can know the number of these windows and bits by calculation. If we only consider $Z_1$-windows and $\mathcal X_1$-windows, the states set here is similar to that in a BB84 protocol\cite{BB84}.
 \\ {\bf Step 4}. {\em They} randomly choose some $Z$-bits to do error test.  By this they can know the bit-error rate in $Z$-basis, $E^Z$.  {\em They} discard the test bits and  the remaining $Z$-bits will be distilled for the final key.
\\
 Note:  For any effective event happens in $Z$-basis, Alice (Bob) judges the bit value in this way: if she (he) has decided to send out a signal pulse, she (he) denotes a bit value 1 (0); if she (he) has decided not to send, she denotes a bit value 0 (1).  One can see straightly, if an effective event happens while both Alice and Bob have decided not to send,  or both of them have decided to send,  a wrong bit in $Z$-basis is created. Because in such a case, the bit value denoted by Alice is different from the bit value denoted by Bob.\\
{\bf Step 5}.  They use the announced data from $X$ pairs to calculate the counting rate (yield) $s_1$ for $\mathcal X_1$-windows (which is also the value for $Z_1$-windows). The number of bits created in $Z_1$-windows can be directly calculated from this value. Also, by observing error rate of $X$ pairs of intensity $\mu$, $E_\mu^X$, the counting rate of intensity $\mu$, $S_\mu$, and the counting rate of vacuum $s_0$, they can calculate the upper bound value of flipping rate of $\mathcal X_1$-bits by
\begin{equation}e_1^{\mathcal X_1}=\frac{S_\mu E_\mu^X-e^{-2\mu}s_0/2}{2\mu e^{-2\mu}s_1}.\label{ph7}
 \end{equation}
 Asymptotically, the phase-flip rate $e_1^{ph}$ for  $Z_1$ bits is $e_1^{ph}=e_1^{\mathcal X_1}$.
\\
Note: In the protocol, Charlie does the beam-splitter measurement\cite{nature} after he takes the phase compensation. There are two output ports of the beam-splitter: right detector and left detector. {\em They} use the following criterion to judge a right bit or a wrong bit in $X$-basis: A right $X$-bit is the left (right) detector clicking caused by an $X$-pair
with positive (negative) value of $\cos(\delta_A-\delta_B)$. A wrong  $X$-bit is the right (left) detector clicking caused by a $X$-pair
with positive (negative) value of $\cos(\delta_A-\delta_B)$.
 Given the observed error rate in $X$-basis and $s_1$,  the phase-flip error rate $e_1^{ph}$ for $Z_1$-bits  can be obtained because asymptotically it is just the error rate of those single-photon-caused $X$-bits, as shown in the supplement. Note that, although {they} know the number of  $\mathcal X_1$-bits, {\em they} don't know which ones are $\mathcal X_1$-bits and hence quantity $e_1^{\mathcal X_1}$ cannot be directly observed, it can be only {\em calculated} by the formula above.
\\
 Note: Also, as one can easily see, if Charlie does the phase compensation perfectly, the out of beam-splitter measurement\cite{nature} will produce a small observed error rate in $X$-basis, if $|\lambda|$ is small in the post-selection criterion Eq.(\ref{slick}). Charlie does not have to be honest or do the compensation perfectly. But this will only change the observed error rate in $X$-basis rather than the security of the protocol.
  \\
{\bf Step 6}. {\em They} distill the final key with an asymptotic  key rate formula
\begin{equation}\label{kr}
N_f = n_1 - n_1 H(e_1^{ph}) - n_t f H(E^Z)
\end{equation}
 $N_f$:   number of final bits, $n_1$:  number  of remaining $Z_1$-bits after error test in Step 4;  $n_t$: number of  remaining $Z$-bits after error test in step 4, $H(x)=-x \log x - (1-x)\log (1-x)$: binary entropy function,  and $f$: error correction efficiency factor.
 The formula can be equivalently written in the following form of key rate per time window:
 \begin{eqnarray}R = 2\epsilon(1-\epsilon)\nonumber\\ \label{krp}
\mu' e^{-\mu'}s_1\left(1 -  H(e_1^{ph})\right) - S_Z f H(E^Z)
\end{eqnarray}
where $S_Z$ is the observed counting rate of $Z$-windows.
\section{Numerical simulation}
In our protocol, we use the traditional formulas for the decoy-state method. Since we don't need any post selection in $Z$ basis and we only need sending or not-sending, there is no misalignment error in this basis. This makes the protocol be able to work with large misalignment from the single-photon interference in $X$ basis. The results of numerical simulation are summarized in Fig.1 and Fig.2.
\begin{figure}%[htb]
    \includegraphics[width=240pt]{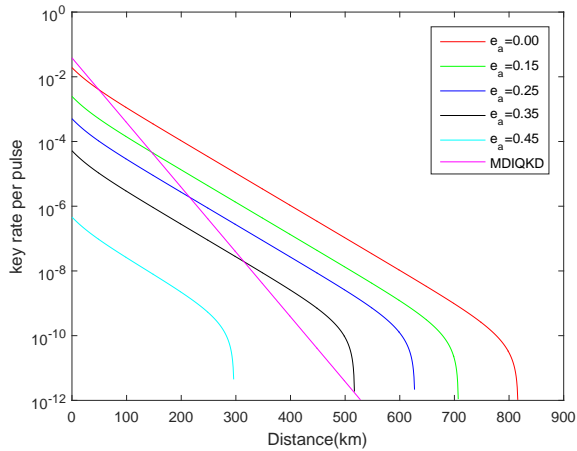}
    \caption{Log scale of the key rate as a function of the distance between Alice and Bob with different misalignment errors. $e_a$: misalignment error rate of single-photon interference. MDIQKD: The optimized key rate for existing decoy-state MDI-QKD with coherent states. In calculating MDI-QKD, we take misalignment error rate $1.5\%$ for $X$-basis and 0 for $Z$-basis.  The numerical result here shows that asymptotically our protocol can have an obvious advantage to the existing decoy-state MDI-QKD even the misalignment error is as large as 35\%.}
\end{figure}
\begin{figure}%[htb]
    \includegraphics[width=240pt]{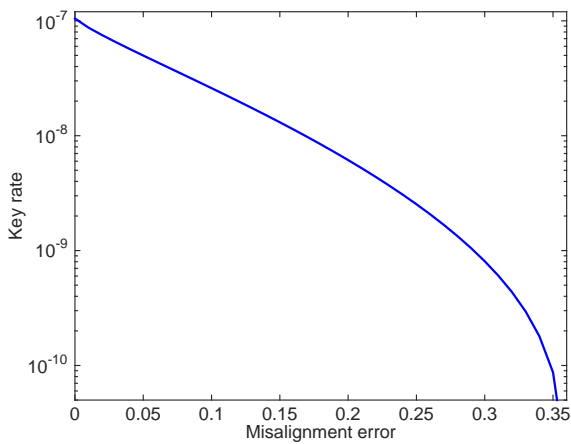}
    \caption{Log scale of the key rate as a function of the misalignment error when the distance between Alice and Bob is 500 km.}
\end{figure}
  In the calculation, we have assumed a detector with dark count rate of $10^{-11}$, and the detection efficiency of $80\%$. An error correction coefficient of 1.1 is set in our calculation. Here, we have only considered the asymptotic result and we have set the phase slice infinitely small. We can do so because in our case we take no post selection in $Z$ basis.  And, at each data, point, we have optimized $\epsilon$ and the signal pulse intensity so as to obtain the best key rate.  We can see that our protocol is so robust to misalignment errors that it can exceed a secure distance of nearly 300 km even with the misalignment error rate of $45\%$. It exceeds a secure distance of 700 km or 600 km even though the single-photon misalignment error rate is as large as $15\%$ or $25\%$.    Also, fixed at the distance to be 500 km, the key rates are shown with different misalignment errors. The largest tolerable error rate can be $35\%$. These results show that our protocol by far breaks the existing a-few-percent  threshold of single-photon misalignment error rate of  for a larger-than-0  secure distance. When there is no misalignment error, our protocol exceeds a secure distance of more than 800 km.\\
\section{Validity of the decoy-state method}
Specifically, in the protocol Alice takes a random phase shift $\delta_A$ to her coherent state  and Bob takes a random phase shift $\delta_B$ to his coherent state. The two-mode weak coherent state prepared by them is $|\sqrt \mu e^{i\gamma_A+i\delta_A}\rangle\otimes |\sqrt \mu e^{i\gamma_B+i\delta_B}\rangle$. Here the global phases $\gamma_A$ and $\gamma_B$ cannot be regarded as random phases because {\em they} also send the strong reference pulses.
 First, we
introduce the new independent variables $\delta_{\pm}=(\delta_B\pm \delta_A)/2$. Integrating the two-mode state of $X$ pulses on variable $\delta_+$ over the range of $[0,2\pi)$, we obtain  a classical mixture  in the convex form
\begin{equation}\sum_k p_k(\mu) |\psi_k\rangle\langle \psi_k|\label{convex}
 \end{equation}with
 $|\psi_k\rangle$ being the state of total photon number $k$ for the two-mode state $|\psi_k\rangle$ and $p_k(\mu)$ being its probability.  For example,
 \begin{equation}\label{ps0}
 |\psi_0\rangle = |00\rangle,\; p_0(\mu) = e^{-2\mu},
 \end{equation}
\begin{equation}\label{ps1}
|\psi_1\rangle = \frac{1}{\sqrt 2}(e^{i\delta_B+i\gamma_B}|01\rangle + e^{i\delta_A+i\gamma_A}|10\rangle)
,
\end{equation}
with
\begin{equation}\label{p1}
p_1(\mu)= 2\mu e^{-2\mu}
\end{equation}
\begin{eqnarray}\label{ps2}
|\psi_2\rangle = \frac{1}{\sqrt 2} e^{i(\delta_A+\gamma_A+\delta_B+\gamma_B)}|11\rangle \nonumber\\+
\frac{1}{2} e^{2i(\delta_B+\gamma_B)}|02\rangle 
 +\frac{1}{2}e^{2i(\delta_A+\gamma_A)}|20\rangle,\nonumber
\end{eqnarray}
with
\begin{equation}
p_2(\mu) = 2\mu^2 e^{-2\mu}\label{p2}
\end{equation}
and so on. This means states from $X$-windows are actually classical mixture of different photon-numbers. The phase randomized states from $Z$-windows can also be regarded as mixture of different photon-number states, in particular, the ingredient of single-photons are randomly on states $|01\rangle$ or $|10\rangle$. As shall be shown later in virtual protocols, single-photon states of Eq.(\ref{p1}) can be used to test the phase-flip rate of those single-photons from $Z$-windows.

One  may argue that there are afterwards announcement of phase information for decoy pulses, how to guarantee the validity of traditional decoy-state method here, e.g., Eq.(\ref{kr}).
Since the phase shift information of signal pulses are never announced, we can regard signal pulses as classical mixture of different photon number states.
 What we want to know is the number of single-photon-caused bits and their phase-flip error rate from signal bits. Once we know the facts,  they do not change by any action outside the lab. Consider a virtual protocol where Alice and Bob secretly decided the random phase shift values prior to the protocol. In such a case, our calculations at Step 6 above is obviously solid. Note that the values of single-photon counts and phase-flip error rate are objective facts which do not change by any outside actions.  After Alice and Bob know the fact, they can announce the phase information of all decoy pulses. But they can also choose to first announce the phase information and then calculate the crucial values for the signal bits, because no one knows at which time they have done the calculation. In such a case, they do not need to predetermine the random phase values, they just use the protocol we proposed above. Also, there is a similar story in  the MDI-QKD: the bases information can not be announced before the states are measured. But it can be announced afterwards, for, the $X$-basis states are only used to know the phase-flip value of those qubits in $Z$ basis.

 Explicitly, we divide the whole space into two subspaces, $\mathcal E$ for Eve and $\mathcal {AB}$ for Alice and Bob. After Alice and Bob post announce phase shift information, they will not receive any information from Charlie (Eve). Suppose Eve has a machine $\mathcal M$ which automatically stores all those post announced information on phase shift values of effective states in $X$-windows.  Eve can in principle have two different choices:
 \\
 Choice 1: Ignores the machine $\mathcal M$ and does not take any actions.
 \\
 Choice 2: Makes use of the stored information of $\mathcal M$ and takes whatever actions she can to her probe.
 \\
 Definitely, under Choice 1, all decoy-state method is valid, all calculated values for signal states such as $\underline {s}_1$ the lower bound of single-photon counts for $Z$-windows and the $\bar {e}_1^{ph}$ upper bound of phase-flip rate of those single-photon counts of $Z$-windows are correct and the final key is secure.
 On the other hand, both Choice 1 and Choice 2 are {\em local} actions in subspace $\mathcal E$ and they do not cause detectable effects in subspace $\mathcal{AB}$.
 Therefore, even Eve takes Choice 2, it makes no difference to subspace $\mathcal AB$. Say, no matter Eve takes which choice, there will be no detectable difference in subspace $\mathcal{AB}$. Therefore, Alice and Bob can always assume Choice 1 for Eve. This can be stated as the following {\bf Theorem}
  \\
  {\bf Theorem} Given whatever information announced by Alice and Bob, Eve's actions to her probe only  cannot cause any detectable effects in Alice and Bob's subspace
$\mathcal {AB}$.

  The phase-flip error is not detectable in the real protocol presented in the earlier section. But, imagine a purification protocol where Alice and Bob use entangled photons in $Z$ windows and coherent states in $X$ windows only. Then the phase-flip error is detectable and the purification result will be all the same no matter which choice Eve has taken. Reducing this virtual protocol to the real protocol we conclude that afterwards announcement of phase shift values does not change the security. Details of this are shown in the  notes of virtual protocol 3.
  \section{Security proof with virtual protocols and reduction}
We first recall the { Definition of time windows} in sending or not-sending (SNS) protocol: Any time window $i$, if both Alice  and Bob commit to a signal window, it is called a $Z$-window; if both of them commit to a decoy window, and if each of them have sent out to Charlie a coherent state of the same intensity $\mu_k$
  it is called an $X$-window. Besides $Z$-windows and $X$-windows, in a complete SNS protocol, there are also mismatching windows, e.g., a time window when Alice commits to a signal window while Bob commits to a decoy window, or, when Alice and Bob each commit to a decoy window but choosing different intensities $\mu_k$ for the coherent state. For presentation conciseness, we shall first prove the security of the {\em simplified form} of  SNS protocol which has $Z$-windows and $X$-windows only. After the {\em simplified} SNS protocol is proven secure, we then show that the proof also holds for the complete SNS protocol.
\subsection{$Z$-basis encoding on ancillary photons of an extended state}
 If the $i$th time window is a $Z$-window, Alice and Bob each  make a decision on either {\em sending} or {\em not-sending}.If Alice (Bob) decides {\em sending},
she (he) puts down a bit value 1 (0) and then sends out a coherent state  to Charlie;   if Alice (Bob) decides {\em not-sending},
she (he) puts down a bit value 0 (1) and does not  send out anything (i.e., sends out a vacuum $|0\rangle$) to Charlie.

 The $Z$-basis encoding of SNS protocol is done by decisions on {\em sending } or {\em not-sending} made by Alice and Bob locally.  More precisely, the {\em sending} or {\em not-sending} decision of a time window that always corresponds to the local classical bits $0$, $1$ to Alice, or 1, 0 to Bob. We can also imagine that whenever Alice (Bob) decides {\em sending} or {\em not-sending}, she (he) always produces a local ancillary photon-number state $|0\rangle$ or $|1\rangle$ and the corresponding bit values are encoded in the local ancillary state. To Alice (Bob), state $|0\rangle$ corresponds to a bit value 0 (1) and state $|1\rangle$ corresponds to a bit value 1 (0). This is equivalent to say that {\em they} (Alice and Bob) have used an extended state including real-photon state which will be sent out to Charlie and ancillary state placed locally. For example, in a certain window when Alice decides sending and Bob decides not sending, we can imagine that {\em they} have actually prepared an extended state
\begin{equation}
\left(\rho_{A}\cdot |0\rangle\langle 0|\right)|\otimes |10\rangle\langle 10| .
\end{equation}
where $\rho_A$ is the coherent state sent out by Alice in a $Z$-window when she decides {\em sending}. We shall also use notation $\rho_B$ as the coherent state sent out by Bob in a $Z$-window when she decides {\em sending}. As stated already, each one's bit value is actually encoded in the local ancillary photon-number state.
If the $i$th time window is a $Z$-window, Alice and Bob each  make a decision on either {\em sending} or {\em not-sending}.
We can also construct an extended quantum state in the complex space $\mathcal T \otimes \mathcal An$ for a $Z$-window as
\begin{align}\label{zt}
&|\Omega\rangle =  (p_1/2) (|0\rangle\langle 0|\cdot\rho_B)|\otimes (|01\rangle\langle 01|\nonumber\\
 &+ (\rho_A\cdot |0\rangle\langle 0)\otimes |10\rangle\langle 10|)\nonumber\\
&+p_2 |00\rangle\langle 00|\otimes |00\rangle\langle 00| \nonumber\\
&+ p_3 (\rho_A\cdot\rho_B)|\otimes |11\rangle\langle 11|
\end{align}
The state $\Omega$ lives in a complex space of $\mathcal T \otimes \mathcal An$. In the right hand side of Eq.(\ref{zt}), those states left to the direct product symbol $\otimes$, such as $|0\rangle\langle 0|\cdot\rho_B$,  $\rho_A\cdot |0\rangle\langle 0$, $|00\rangle\langle 00|$, and
$\rho_A\cdot\rho_B$ are in the subspace $\mathcal T$ and those states right to the direct product $\otimes$, such as
$|10\rangle\langle 10|$,  $|01\rangle\langle 01|$, $|00\rangle\langle 00|$, and $|11\rangle\langle 11|$ are in the subspace $\mathcal An$.
For presentation simplicity, we shall name the light field of subspace $\mathcal T$ in an extended state as {\em real-photon state},  or {\em real photons}, and  name the local light field in subspace $\mathcal An$ as {\em ancillary-photon state},  or {\em ancillary photons}.

Ancillary state $|01\rangle\langle 01|$ ($|10\rangle\langle 10 |$)
is for the decisions that Alice decides {\em not-sending} ({\em sending}) and Bob decides {\em sending} ({\em not-sending}). Ancillary-photon state $|00\rangle\langle 00|$ ($|11\rangle\langle 11|$) is for the decisions that both of them decide {\em not-sending} ({\em sending}). In a $Z$-window of SNS protocol, {\em their} action is equivalent to  just sending out the {\em real photons} of $\Omega$ to Charlie and keep their {\em ancillary photons}.

Also, since $\rho_A$ and $\rho_B$  are phase-randomized coherent states, each of these states  can be regarded as classical mixtures of different photon number states. Say, we can replace $\rho_A$ or  $\rho_B$ by
 \begin{equation}\label{con}
 \rho_{\mu'} = \sum_{n=0}^\infty \frac{e^{-\mu'}\mu'^n}{n!}|n\rangle\langle n|= \mu'e^{-\mu'}|1\rangle \langle 1| + (1-\mu'\ e^{-\mu'})\bar\rho
 \end{equation}
 where
 \begin{equation}\label{br}
 \bar\rho = \frac{1}{1-\mu' e^{-\mu'}}\sum_{n\not=1} \frac{e^{-\mu'}\mu'^n}{n!}|n\rangle\langle n|
 \end{equation}
 and hence we can rewrite the extended state $\Omega$ in the following equivalent format:
 \begin{equation}\label{zt0}
\Omega =  \sum_r q_r\Omega_r
\end{equation}
where $r=1,2,3,4$ and
 \begin{align}\label{zt1}
&\Omega_1 =  (1/2) (|01\rangle\langle 01|\otimes |01\rangle\langle 01| \nonumber\\
&+ |10\rangle\langle  10|\otimes |10\rangle\langle 10|)\nonumber\\
&\Omega_2 = (|0\rangle\langle 0|\cdot \bar \rho) \otimes |01\rangle\langle 01|\nonumber\\
 &+ (\bar\rho\cdot |0\rangle\langle 0|)\otimes  |10\rangle\langle 10|\nonumber\\
&\Omega_3=|00\rangle\langle 00|\otimes |00\rangle\langle 00|\nonumber\\
&\Omega_4=\rho_{\mu'} \cdot \rho_{\mu'} | \otimes |11\rangle \langle 11|
\end{align}
 Also, for any time window $i$, if it is an $X$-window of SNS protocol, {\em they} (Alice and Bob) send out two-mode coherent state
 \begin{equation}\label{xx}
 \rho_{X}= |\tilde{\beta}_k\rangle\langle\tilde {\beta}_k|
 \end{equation}
 where , i.e., in the form of a two-mode coherent state
\begin{equation}\label{beta}|\tilde{\beta}_k\rangle=|\sqrt\mu_k e^{i\delta_{A}+i\gamma_{A}}\rangle|\sqrt\mu_k e^{i\delta_{B}+i\gamma_{B}}\rangle
,\end{equation} and $k$ is randomly chosen from
 a few different values for different intensities $\mu_k$, $\delta_{A},\;\delta_{B}$ are random values taken privately by Alice and Bob, respectively, $\gamma_{A},\gamma_{B}$ are global phases announced to Charlie publicly.

 In the  SNS protocol above, the state for a $Z$-window is a classical mixture of different different kinds of time windows. The $Z$-windows are classical mixture of $Z_1$-windows which only uses the extended states $\Omega_1$ and other types of $Z$-windows which uses the extended states of $\Omega_2,\Omega_3,\Omega_4$. (Note that
 all these states are orthogonal.)
 To show the security of this  protocol, we can take the following theme:
 We first show the security of a protocol with only state $\Omega_1$ for a $Z$-window, and then extend it to the case of state $\Omega$ for a $Z$-window by the tagged model\cite{gllp}: We regard the bit values of $Z$-basis encoding from state $\Omega_1$ as the set of {\em un-tagged bits} and the bit values from other states
 ($\Omega_2,\Omega_3,\Omega_4$) as the set of {\em tagged bits}.

 In a complete SNS protocol, besides $X$-windows and $Z$-windows, there are other time windows (those mismatching windows\cite{note1}), but as shown in the end of the proof, in that case another extended state including all time windows is constructible and it is still a mixture of $\Omega_1$ and other states therefore the tagged model and the security proof here still holds. At this moment, for presentation conciseness, we  consider the {\em simplified form} of SNS protocol where there are only $Z$-windows and $X$-windows.
 \subsection{Virtual protocol 1}
 \noindent
 {\bf Definition} of {\em effective event}: We define an {\bf effective event} of a $Z$-window if Charlie announces one and only one detector clicking for an individual  $Z$-window.  We define an {\bf effective event} of an $X$-window if Charlie announces one and only one detector clicking for an individual $X$-window and values $\delta_A,\delta_B$ in the corresponding state satisfies Eq.(\ref{slice}). {\em They} will then only use states or data corresponding to  effective events in the protocol. A time window that presents an effective event is named as an effective time window. An {\em effective ancillary photon} is an ancillary
  photon corresponding to an effective event.
 \\{\em Preparation stage}\\
 {\em They} pre-share classical information for different time windows {\em they} will use, $X$-windows and $Z$-windows. {\em They} also pre-share
  an extended state
\begin{align}\label{om12}
&\Omega_{0i} = |\Psi_{1i}\rangle\langle\Psi_{1i}|\nonumber\\
& |\Psi_{1i}\rangle = e^{i\gamma_{B_i}}|01\rangle\otimes|01\rangle+e^{i\gamma_{A_i}}|10\rangle\otimes|10\rangle
%& 1-|\cos(\delta_{B_i}-\delta_{A_i})| \le |\lambda|\label{cons}
\end{align}
for the $i$th time window.
Here  values of $\gamma_{A_i},\gamma_{B_i}$ are  announced publicly.

For any time window $i$, if it is an $X$-window,  Alice takes a local random phase shift $\delta_{Ai}$ and
Bob takes a local random phase shift $\delta_{Bi}$ locally to the real-photon of state $\Omega_{0i}$.
We name the state after the random phase shifts as $\Omega_{Xi}$.  Explicitly
\begin{align}\label{omx}
&\Omega_{X_i} = |\Psi_{1i}'\rangle\langle\Psi_{1i}'|\nonumber\\
& |\Psi_{1i}'\rangle = e^{i\delta_{B_i}+i\gamma_{B_i}}|01\rangle\otimes|01\rangle+e^{i\delta_{A_i}+i\gamma_{A_i}}|10\rangle\otimes|10\rangle
\end{align}
with the random values $\delta_{Ai},\delta_{Bi}$ being privately chosen by Alice and Bob, respectively.

For any time window $i$, if it is an $Z$-window,  through discussions by a secret channel, Alice takes a local restricted random phase shift $\delta_{Ai}$ and
Bob takes a local restricted random phase shift $\delta_{Bi}$  to the real-photon of state $\Omega_{0i}$, with the restriction
\begin{equation}
1-|\cos(\delta_{B_i}-\delta_{A_i})| \le |\lambda|\label{cons}
\end{equation}
We name the state after the restricted random phase shifts as $\Omega_{Zi}$, which has the form
\begin{align}\label{omz}
&\Omega_{Z_i} = |\Psi_1'\rangle\langle\Psi_{1i}'|\nonumber\\
& |\Psi_{1i}'\rangle = e^{i\delta_{B_i}+i\gamma_{B_i}}|01\rangle\otimes|01\rangle+e^{i\delta_{A_i}+i\gamma_{A_i}}|10\rangle\otimes|10\rangle
\end{align}
which $\Omega_{Xi}$, but with an additional restriction
of Eq.(\ref{cons}).

The constraint Eq.(\ref{cons}) makes the state in $Z$-windows  not identical to that in all $X$-windows.
If we define an $\tilde X$-window as a time window whose parameters $\delta_{A_i}, \delta_{B_i}$ in extended state $\Omega_{X_i}$ satisfying Eq.(\ref{cons}),
the extended state for $Z$-windows is identical to the extended state of $\tilde X$-windows.

For presentation simplicity, we shall omit the subscripts $i$ in all phase values $\delta_{A_i},\delta_{B_i},\gamma_{A_i},\gamma_{B_i}$ and states.
Also, we introduce states $|\chi^0\rangle, |\chi^1\rangle$ in the real-photon space for any time window
 \begin{align}
  & |\chi^0\rangle = \frac{1}{\sqrt 2}(e^{i\delta_B+i\gamma_B}|01\rangle+e^{i\delta_A+i\gamma_A}|10\rangle)\nonumber\\
  & |\chi^1\rangle = \frac{1}{\sqrt 2}(e^{i\delta_B+i\gamma_B}|01\rangle-e^{i\delta_A+i\gamma_A}|10\rangle)\nonumber\\
  & {\rm if}\; \cos(\delta_B-\delta_A) \ge 0  \label{chia}
  \end{align}
  and
  \begin{align}
  & |\chi^0\rangle = \frac{1}{\sqrt 2}(e^{i\delta_B+i\gamma_B}|01\rangle-e^{i\delta_A+i\gamma_A}|10\rangle)\nonumber\\
  & |\chi^1\rangle = \frac{1}{\sqrt 2}(e^{i\delta_B+i\gamma_B}|01\rangle+e^{i\delta_A+i\gamma_A}|10\rangle)\nonumber\\
  & {\rm if}\; \cos(\delta_B-\delta_A) < 0.\label{chib}
  \end{align}
\\{\em Virtual Protocol} 1\\
1-1 At any time window $i$, if it is a $Z$-window ($X$-window), {\em they} send out to Charlie the real-photon state from state $\Omega_Z$ ($\Omega_X$) as defined
by Eq.(\ref{omz}) (Eq.(\ref{omx}))to Charlie
 and keep the ancillary photons locally.
\\1-2 Charlie announces his measurement outcome of  all time windows.  {\em They} tell each other $\delta_A,\delta_B$ values through  classical communication  and then take post selection to all $X$-windows and the one-detector-clicking events from the $X$-windows by the following criterion
    \begin{equation}
    \label{slice}
    1- |\cos(\delta_B-\delta_A)|\le |\lambda|
    \end{equation}
    which is identical to Eq.(\ref{cons}). Taking post selection by this criterion, {\em they} obtain $\tilde X$-windows and effective events of $X$-windows which can be regarded as effective events of $\tilde X$-windows.
     According to our definition, an $\tilde X$-window satisfies Eq.(\ref{cons}) therefore identical to a $Z$-window.
\\{\em Definition}: After the post selection taken in step 1-2, {\em they} divide their effective time windows and corresponding effective ancillary photons into 4 subsets according to the
clicking detector (the left or the right) and the sign of $\cos(\delta_B-\delta_A)$ (positive or negative).  Each subset of time windows is labeled by $\xi=(a,d)$ where
$a=+,-$ and $d=L,R$.
\\Explicitly,  time window $\xi=(a,d)$ is an effective time window heralded by joint events of $a$ and $d$ as defined in the following:
\\Event $a$: the sign of $\cos(\delta_B-\delta_A)$ is $a$ ($+$ or $-$). Explicitly, $a=+$ for $\cos(\delta_B-\delta_A)\ge 0$, $a=-$ for $\cos(\delta_B-\delta_A)< 0$.
\\Event $d$: Detector $d$ has clicked and the other detector has not clicked. $d$ can be either $L$ for the left detector or $R$ for the right detector.
\\ \underline {Definitions} We shall use notation $Z_{\xi}$ ($X_{\xi}$) for a $Z$-window ($X$-window) with joint events  of $a$, $d$ for $\xi=(a,d)$. We shall also
use set $\mathcal A_{Z_{\xi}}$ ($\mathcal A_{X_{\xi}}$) for the set of effective ancillary photons of time windows $Z_{\xi}$ ($X_{\xi}$).
 \\1-3  {\em They} check the phase-flip error rate $E_{\xi}$ for set of $\mathcal A_{ X_{\xi}}$, where $\xi=(+,L),(-,L),(+,R),(-,R)$, which is also the estimated phase-flip error rates of set $\mathcal A_{ Z_{\xi}}$ and $\xi=(+,L),(-,L),(+,R),(-,R)$.
 \\1-4 {\em They }  purify the  ancillary photons of time windows $ Z_{\xi}$ and $\xi=(+,L),(-,L),(+,R),(-,R)$  separately. After purification, {\em they} obtain high quality single-photon states $|\Phi^0\rangle$ or $|\Phi^1\rangle$ with (almost) 100\% purity.  {\em They} each measures the photon number locally to the purified photons and obtain the final key $k_f$. Alice puts down a bit value 0 or 1 whenever she obtains a measurement outcome of vacuum or 1 photon, Bob puts down a bit value 1 or 0 whenever she obtains a measurement outcome of vacuum or 1 photon.
\\ \underline {Note 1} {\em Security }. The security of the final key is based on the faithfulness of the purification, i.e., the estimation of phase-flip error rate. Charlie has determined  effective ancillary photons but Alice and Bob test the phase-flip error rate themselves in step 1-3.
 Although the extended state of an $X$-window is not identical to that of a $Z$-window, the extended state
of an $\tilde X$-window {\em is} identical to that of a $Z$-window. After the post-selection condition in step 1-2, it is equivalent to say that all effective events of $X$-windows are just effective events from $\tilde X$ windows. Therefore, an ancillary photon from set  $\mathcal A_{X_{\xi}}$ is identical to an ancillary photon from set $\mathcal A_{Z_{\xi}}$.  So, statistically, the phase-flip-error rate value of set $\mathcal A_{ X_{\xi}}$ is exactly the  value of set $\mathcal A_{ Z_{\xi}}$.
\\ \underline{ Note 2} { Definitions of phase-flip-error rate}. \\
   % {\em Definition 1} for phase-flip-error rate.
  Suppose set $\mathcal A_{X_{\xi}}$ contains $n_{\xi}$ effective ancillary photons. If each photons of set $\mathcal A_{X_{\xi}}$ were measured in basis $\{|\Phi^0\rangle, |\Phi^1\rangle\}$ and there were
   $n_{\xi}^{(0)}$ outcome of $|\Phi^0\rangle\langle\Phi^0|$, and $n_{\xi}^{(1)}$ outcome of $|\Phi^1\rangle\langle\Phi^1|$, the phase-flip error rate for set $\mathcal A_{ X_{\xi}}$ is
   \begin{equation}\label{er}
   E_{\xi}= \frac{{ \min}\left(n_{\xi}^{(0)},n^{(1)}_{\xi}\right)}{n_{\xi}}.
   \end{equation}
   Changing the values of $n_{\xi}^{(0)},n^{(1)}_{\xi},n_{\xi}$ into the corresponding values of set $\mathcal A_{Z_{\xi}}$ in Eq.(\ref{er}), we can define the phase flip error rate for set $\mathcal A_{Z_{\xi}}$. Statistically,  $E_{\xi}$ for set
   $\mathcal A_{X_{\xi}}$  is also
    the asymptotic phase-flip error rate of set $\mathcal A_{ Z_{\xi}}$.
    To know
   the values $E_{\xi}$, {\em they} can choose to measure each photons of set $\mathcal A_{X_{\xi}}$ in basis $\{|\Phi^0\rangle, |\Phi^1\rangle\}$. But instead of this, {\em they} can also choose to take  local measurements in basis $\{|x\pm\rangle\}$ in each sides and check the parity of each  measurement outcome. (Outcome of $|x+\rangle|x+\rangle$ or
    $|x-\rangle|x-\rangle$) are even-parity while $|x+\rangle|x-\rangle$ or $|x-\rangle|x+\rangle$ are odd parity.) Note that all effective ancillary photons are single-photons. As it is easy to see , for single-photons, the fraction of odd parity (even parity) outcome from  measurement of each sides in basis $\{|x\pm\rangle\}$
   is exactly equal to the fraction of $|\Phi^1\rangle\langle\Phi^1|$ ($|\Phi^0\rangle\langle\Phi^0|$) outcome from the measurement in basis  $\{|\Phi^0\rangle, |\Phi^1\rangle\}$. Moreover, this measurement step is only needed here for this Virtual protocol, it is not needed for a real protocol. For ease of presentation,
   we suppose {\em they} use the measurement basis $\{|\Phi^0\rangle,|\Phi^1\rangle\}$.
  \\ \underline{Note 3} { Reduction of pre-shared states for $X$-windows}\\
   \underline{{\em Reduction 1}}
  It makes no difference to anyone outside  if {\em they} measure all ancillary photons of $X$-windows in basis $\{|\Phi^0\rangle,|\Phi^1\rangle\}$
   before the protocol starts. This measurement operation is on ancillary photon while the initial random phase shift operation ($\delta_A,\delta_B$) are on the real-photon space, so these two operation commute. We assume {\em they} first take measurement to ancillary photons and then take local random phase shifts to the real-photon state for an $X$-window. {\em They} start from the pre-shared pair of Eq.(\ref{om12}). After measurement to the ancillary photon, {\em they} obtain one of the following outcome extended state  for an $X$-window, depending on the measurement outcome of ancillary photon:\\
   either
   \begin{align}\label{s1}
   & |\tilde W_0\rangle\otimes |\Phi^0\rangle\nonumber\\
   & |\tilde W_0\rangle=\frac{1}{\sqrt 2}(e^{i\gamma_B}|01\rangle+ e^{i\gamma_A}|10\rangle)
    \end{align}
    or
    \begin{align}\label{s2}
   & |\tilde W_1\rangle\otimes |\Phi^1\rangle\nonumber\\
   & |\tilde W_1\rangle=\frac{1}{\sqrt 2}(e^{i\gamma_B}|01\rangle - e^{i\gamma_A}|10\rangle)
    \end{align}
    {\em They} then take local phase shifts $\delta_A,\delta_B$ to real-photon state of outcome extended state, which is one of the above two states.
    If {\em They} then take all steps in Virtual 1 from step 1-1 to step 1-4 as if {\em they} were using the original pre-shared extended states without measurement to the ancillary
    photons at this stage. The result should be equivalent to the original Virtual protocol.
    \\ \underline{{\em Reduction 2}}
    Alternatively, {\em they} can just start with states of Eq.(\ref{s1},\ref{s2}) for their $X$-windows. {\em They} need pre-share classical information on $Z$-windows, $X_0$-windows, and $X_1$-windows. {\em They}  pre-share real-photon states $|\tilde W_0\rangle=\frac{1}{\sqrt 2}(e^{i\gamma_B}|01\rangle+ e^{i\gamma_A}|10\rangle)$ for $X_0$-windows and $|\tilde W_1\rangle=\frac{1}{\sqrt 2}(e^{i\gamma_B}|01\rangle- e^{i\gamma_A}|10\rangle)$ for $X_1$-windows. Imagine that {\em they} also pre-share some single-photon states $|\Phi^0\rangle$ and $|\Phi^1\rangle$. (These states $|\Phi^0\rangle$ and $|\Phi^1\rangle$ are not really necessary, to show everything clearly we assume so at this moment.)
\\In an $X_0$-window, {\em they} take local private random phase-shift $\delta_A,\delta_B$ on the pre-shared state $|\tilde W_0\rangle$, changing it to
  \begin{equation}
  |W_0\rangle = \frac{1}{\sqrt 2}(e^{i\delta_B+i\gamma_B}|01\rangle + e^{i\delta_A+i\gamma_A}|10\rangle).
  \end{equation}
  {\em They} label a pre-shared state $|\Phi^0\rangle$ as the ancillary photon for this state $|W_0\rangle$ above.
  {\em They} then send the real-photon state $|W_0\rangle$  out to Charlie.
  After step 1-2, {\em they} have known the values of $\delta_A,\delta_B$,  and {\em they} now know the original extended state with the labeled ancillary photon
  \begin{align}\label{kaka1}
 & \Omega_{+,0} = |\chi^0\rangle\langle \chi^0| \otimes |\Phi^0\rangle\langle\Phi^0|\;{\rm if}\; \cos(\delta_B-\delta_A)\ge 0\\
 & \Omega_{-,0} = |\chi^1\rangle\langle \chi^1| \otimes |\Phi^0\rangle\langle\Phi^0|\;{\rm if}\; \cos(\delta_B-\delta_A)< 0\label{kaka2}
  \end{align}
  Here we have used the same definition for $|\chi^0\rangle,|\chi^1\rangle$ as used in Eq.(\ref{chia},\ref{chib}).
  \\ In an $X_1$-window, {\em they} take the same operations above to state $|\tilde W_1\rangle$, changing it to
  \begin{equation}
  |W_1\rangle = \frac{1}{\sqrt 2}(e^{i\delta_B+i\gamma_B}|01\rangle - e^{i\delta_A+i\gamma_A}|10\rangle).
  \end{equation}
  {\em They} label a pre-shared state $|\Phi^1\rangle$ as the ancillary photon for this state $|W_1\rangle$ above.
  {\em They} then send the real-photon state $|W_1\rangle$  out to Charlie.
  After step 1-2, {\em they} will know the values of $\delta_A,\delta_B$,  and {\em they} now know the original extended state with the labeled ancillary photon is
  \begin{align}
 & \Omega_{+,1} = |\chi^1\rangle\langle \chi^1| \otimes |\Phi^1\rangle\langle\Phi^1|\;{\rm if}\; \cos(\delta_B-\delta_A)\ge 0\label{kaka3}\\
 & \Omega_{-,1} = |\chi^0\rangle\langle \chi^0| \otimes |\Phi^1\rangle\langle\Phi^1|\;{\rm if}\; \cos(\delta_B-\delta_A)< 0\label{kaka4}
  \end{align}
  Here we have used the same definition for $|\chi^0\rangle,|\chi^1\rangle$ as used in Eq.(\ref{chia},\ref{chib}).
 \\
 Given the orthogonal extended states by Eqs(\ref{kaka1},\ref{kaka2},\ref{kaka3},\ref{kaka4}),  we can  define
   4 subsets  of time windows by $X_{(a,b)}$, where $a=+,-$ and $b=0,1$. An $X_{(a,b)}$-window is an effective time window  heralded by joint events $a$ and $b$ defined in the following:
   \\ Event $a$: the sign of $\cos(\delta_B-\delta_A)$ is $a$;
   \\ Event $b$: the ancillary state is $|\Phi^b\rangle$. Specifically,
  \begin{align}\label{def}
  & X_{(a,b)}-{\rm window}:\;\nonumber\\
  & a=+\; {\rm for}\; \cos(\delta_B-\delta_A)\ge 0\nonumber\\
  & a=-\; {\rm for}\; \cos(\delta_B-\delta_A)< 0\nonumber\\
  & b=0\; {\rm for}\; {\rm ancillary}\;{\rm state}\; |\Phi^0\rangle\langle \Phi^0|\nonumber\\
  & b=1\; {\rm for}\; {\rm ancillary}\;{\rm state}\; |\Phi^1\rangle\langle \Phi^1|.
  \end{align}
  On the other hand, after step 1-2, {\em they} can judge explicitly the values $a$ and $b$ if it is an effective window. Value $b$ is determined by the pre-shared information,
  $b=0$ for an $X_0$-window and $b=1$ for an $X_1$-window. Value $a$ is determined by the random phase shift values of $\delta_A,\delta_B$ chosen for the time window,
  $a=+$ if  $\cos(\delta_B-\delta_A)\ge 0$, $a=-$ if $\cos(\delta_B-\delta_A)<0$.
     \\Given an $X_0$-window or an $X_1$-window, the measurement outcome in basis $\{|\Phi^0\rangle,|\Phi^1\rangle\}$ in step 1-3 is actually deterministic and hence the measurement in step 1-3 is not necessary.
     Therefore, according to our {\em  Definition  1},  {\em they} can use  the following operable definition to calculate each quantities in Eq.(\ref{er})
     after step 1-2. We introduce $X_{(a,b,d)}$ for an effective time window with joint events $a,\;b,$ and $d$, as defined in the following:
      \\ Event $a$: The sign of $\cos(\delta_A-\delta_B)$;
      \\Event $b$: The time window is $X_b$-window;
      \\Event $d$: Detector $d$ has clicked and the other detector has not clicked, $d=L$ for left detector and $d=R$ for the right detector.
     \\For example an $X_{(+,1,L)}$-window is a time window satisfying the following conditions:
      \\1, At this window, $\cos(\delta_B-\delta_A)\ge 0$
      \\ 2, It is an $X_1$-window, i.e. the the ancillary photon state is $|\Phi^1\rangle\langle\Phi^1|$
     \\ 3, The left detector clicks and the right detector does not click.\\
     We also introduce notation $N_{X_{(a,b,d)}}$ for the number of $X_{(a,b,d)}$-windows in the protocol. Therefore we have
     \begin{align}
     & n_{(a,d)}^{(0)} = N_{X_{(a,0,d)}}\label{er1}\\
     & n_{(a,d)}^{(1)} = N_{X_{(a,1,d)}}\label{er2}
     \end{align}
     for Eq.(\ref{er}).
     Given Eq(\ref{er1},\ref{er2}), we can apply Eq.(\ref{er}) immediately after step 1-2, i.e., we have removed the measurement operation in step 1-3.
     \\ {\em Importantly}, all values of $a,b,c$ can be determined from the values of $\delta_A,\delta_B$, the pre-shared information for time window $X_0$ or
      $X_1$, and Charlie's announcement on the clicking detector, $L$ or $R$. The ancillary photons for $X$-windows  are actually {\em not} needed in the protocol.
\subsection{Virtual protocol 2}
\noindent Here we assume {\em they} pre-share a classical information for windows of $Z$, $X_0$, and $X_1$. {\em They } pre-share the same extended states $\Omega_Z$ for $Z$-windows as in Virtual protocol 1.
{\em They} initially pre-share  real-photon states $|\tilde W_0\rangle=\frac{1}{\sqrt 2}(e^{i\gamma_B}|01\rangle+ e^{i\gamma_A}|10\rangle)$ for $X_0$-windows and $|\tilde W_1\rangle=\frac{1}{\sqrt 2}(e^{i\gamma_B}|01\rangle- e^{i\gamma_A}|10\rangle)$ for $X_1$-windows. {\em They} take local random phase-shifts $\delta_A,\delta_B$ on a state $|\tilde W_0\rangle$ for an $X_0$-window, on state $|\tilde W_1\rangle$ for an $X_1$-window. After local phase-shifts, {\em they} share a state
$|W_0\rangle = \frac{1}{\sqrt 2}(e^{i\delta_B+i\gamma_B}|01\rangle + e^{i\delta_A+i\gamma_A}|10\rangle)$ for an $X_0$-window and a state $|W_1\rangle = \frac{1}{\sqrt 2}(e^{i\delta_B+i\gamma_B}|01\rangle - e^{i\delta_A+i\gamma_A}|10\rangle)$ for  an $X_1$-window.
\\{\em Virtual Protocol} 2\\
2-1 At any time window $i$, if it is a $Z$-window, {\em they} send out to Charlie the real-photon  from state $\Omega_Z$  to Charlie
 and keep the ancillary photon locally. If  it is an $X_0$-window ($X_1$-window), {\em they} send out to Charlie the real-photon state
 $|W_0\rangle$ ($|W_1\rangle$).
 \\2-2 Charlie announces his measurement outcome of  all time windows.  {\em They} tell each other $\delta_A,\delta_B$ values through classical communication  and then take post selection for   $X$-windows  by criterion of Eq.(\ref{slice}).
    \\2-3  {\em They} estimate the phase-flip error rate $E_{\xi}$ for sets of $\mathcal A_{Z_{\xi}}$, where $\xi=(+,L),(-,L),(+,R),(-,R)$ by formula
    \begin{equation}\label{pha2}
    E_{(a,d)} = \frac{\min (N_{X_{(a,0,d)}},N_{X_{( a,1,d)}})}{n_{{(a,d)}}}
    \end{equation}
    where $d=L,R$.
    % $\bar a$ is the different binary value of $a$.
 \\2-4 {\em They }  purify the  effective ancillary photons in sets $\mathcal A_{Z_{\xi}}$ and $\xi=(+,L),(-,L),(+,R),(-,R)$  separately. After purification, {\em they} obtain a number of final states all
 in $|\Phi^0\rangle$ from sets $(+,L),(-,R)$, and all in  $|\Phi^1\rangle$ from sets $(-,L),(+,R)$.  {\em They} each measures the photon-number locally to each  purified single-photons and obtain the final key $k_f$.
 \\
 \underline{Note 1} The $X_1$-window is not needed. It is easy to show, the  density operator $\rho_0$ for a time window $X_0$ is actually identical to the density operator $\rho_1$
 for a time window $X_1$. Also, it is easy to see
 \begin{equation}
 \rho_{+,0}=\rho_{-,1},\; \rho_{-,0}=\rho_{+,1}
 \end{equation}
 where $\rho_{a,b}$ is the density operator for time windows of $X_{(a,b)}$, taken average on all allowed values of $\delta_A,\delta_B$. This means we have
 \begin{equation}\label{subs}
 N_{X_{(a,1,d)}}=N_{X_{(\bar a,0,d)}}
 \end{equation}
 therefore we can simply replace
 $N_{X_{(a,1,d)}}$ in the phase-flip error rate formula Eq.(\ref{pha2}) by $N_{X_{(\bar a,0,d)}}$. Also since $\rho_0=\rho_1$, Eve can find no difference if we replace all $X_1$-windows by $X_0$-windows. Therefore, we don't need $X_1$-windows, consequently, {\em they} only need a classical information for $Z$-windows and $X$-windows (i.e., $X_0$-windows), and {\em they} only need an initial state $|\tilde W_0\rangle$ for $X$-windows. In this way, an $X$-window is just an $X_0$-window. Consider
 $N_{X_{a,1,d}}$ in  Eq.(\ref{pha2}). It can be replaced by $N_{X_{\bar a, 0, d}}$ because of Eq.(\ref{subs}). Further, since there is no $X_1$-window now, $X_0$-window is just $X$-window,  $N_{X_{a,1,d}}$ can be further replaced by $N_{X_{( \bar a,d)}}$ and Eq.(\ref{pha2}) is replaced by
 \begin{equation}
  E_{(a,d)} = \frac{\min (N_{X_{(a,d)}},N_{X_{( \bar a,d)}})}{n_{{(a,d)}}} \label{pha3}
 \end{equation}
 \underline{Note 2} {\em They} don't need to pre-share any state for $X$-windows. As was shown by Eq.(\ref{convex}) already, the two-mode coherent state  can be regarded as a mixture of different two-mode photon number state. The single-photon state there in Eq.(\ref{ps1}) is equivalent to the pre-shared state of $| W_0\rangle$.
 \\
 \underline{Note 3} Purifying all effective ancillary photon in one batch. Definitely, {\em they} can choose to purify all effective
 ancillary photons of $Z$-windows in one batch. The phase-flip error rate is
 \begin{align}\label{phase1}
 & E^{ph} = \frac{\sum_{a,d}\min(N_{X_{(a,d)}},N_{X_{(\bar a,d)}})}{n_1}\\
 & = \frac{ 2\sum_d{\min(N_{X_{(+,d)}},N_{X_{( -,d)}})}}{N_{X_{(+,L)}}+N_{X_{(-,L)}}+N_{X_{(+,R)}}+N_{X_{(-,R)}}}\label{phat}
 \end{align}
 where $n_1=N_{X_{(+,L)}}+N_{X_{(-,L)}}+N_{X_{(+,R)}}+N_{X_{(-,R)}}$ is the total number of effective  $X$-windows.
 Surely, $N_{X_{(-,L)}}\ge \min(N_{X_{(+,L)}},N_{X_{( -,L)}})$ and $N_{X_{(+,R)}}\ge \min(N_{X_{(+,R)}},N_{X_{( -,R)}})$.
  Therefore the phase-flip error rate formula of Eq.(\ref{phase1}) can be simplified into
 \begin{equation}\label{phase2}
 E^{ph} \le \frac{N_{X_{(-,L)}}+N_{X_{(+,R)}}}{n_1}
 \end{equation}
 which is simply to count the following two types of joint events as phase-flip errors:
 \\ 1, Left-detector-clicking only and $\cos(\delta_B-\delta_A)<0$
 \\ 2, Right-detector-clicking only and $\cos(\delta_B-\delta_A)\ge 0$.
 \\ If {\em they} use this formula, Charlie can make a high quality raw state of effective ancillary photons for Alice and Bob by setting his measurement set-up
 properly so that
 with very small probability for the left-detector-clicking (right-detector-clicking) due to the incident state of $|\chi^1\rangle$ ($|\chi^0\rangle$).
  \subsection{Virtual protocol 3}
\noindent 3-1 {\em They} send out the real photons of state $\Omega_Z$ in Eq.(\ref{om12}) for a $Z$-window and state $\rho_X$ as defined in Eq.(\ref{xx}) in an $X$-window.\\
3-2,   Charlie announces his measurement outcome. {\em They} each announce the random phase shift values $\delta_A,\delta_B$ and take post selection for  $X$-windows by Eq.(\ref{slice}).
\\3-3 {\em They} verify the phase-flip error rate $e_1^{ph}$ for effective ancillary photons with classical data of $X$-windows announced by Charlie through decoy-state analysis. In an $X$-window, an error is counted if  the $\cos(\delta_B-\delta_A)\ge 0$ and right detector clicks,
 or $\cos(\delta_B-\delta_A)<0$ and the left detector clicks.\\
3-4 {\em They} take purification and local measurement on purified single-photons to obtain the final key.
\\ \underline{Note 1} $e_1^{ph}$, $E^{ph}$, and {\em validity of the decoy-state method}
 \\The physical meaning of $e_1^{ph}$ is same with that of $E^{ph}$ that appeared in  Virtual protocol 2, just  the phase-flip error rate of effective ancillary photons of $Z$-windows. But there, the value $E^{ph}$ is directly observed, here the value $e_1^{ph}$ is calculated by the decoy-state method.
 \\We use notation ${\mathcal I}$ for the information of random phase-shift values $\delta_A,\delta_B$ of state $\rho_X$ post  announced in step 3-2.
     According to our {\bf Theorem}, Eve's action
    with information $\mathcal I$ does not cause any detectable effects  for any set of ancillary photons.
    Therefore, any physically testable conclusion on the ancillary photons, if it is correct in the case that Eve ignores information $\mathcal I$,
    it must be also correct in the case that Eve uses $\mathcal I$.
 Here the decoy-state analysis is to conclude  the upper bound value of phase-flip error rate of the effective ancillary photons. The conclusion is
 physically testable because the phase-flip error rate for the ancillary photons here is physically detectable.
 Definitely, the conclusion form the decoy-state method for the upper bound is correct if Eve ignores information $\mathcal I$. According to our {\bf Theorem} in Section {\bf IV}
 the upper-bound conclusion must be also correct in the case that Eve uses $\mathcal I$.
\\ \underline{Note  2}: {\em Probabilistic mixture of different photon-number states and the decoy-state analysis}
 \\ Consider Eqs(\ref{convex},\ref{ps1}). We can regard the $X$-windows as classical mixture of $\mathcal X_1$-window and other types of $X$-windows, and an $\mathcal X_1$-window is defined as an $X$-window when a two-mode single-photon is sent out to Charlie by Alice and Bob.
 We need the yield value of $s_1$, which is just the effective-event rate of all $\mathcal X_1$-windows. Say, $k_1$ effective events are produced from $K_1$ $\mathcal X_1$-windows in the whole protocol, then $s_1=k_1/K_1$.
 This can be worked out by decoy-state analysis, e.g., given 3 intensities $\mu_0=0,\mu_1,\mu_2$ and $\mu_0=0<\mu_1<\mu_2$, through directly applying Eq.(17) of Ref\cite{decg} we have:
 \begin{align}& s_1\ge \underline{s}_1=\nonumber
 \\
  & \frac{p_2(\mu_2)(S_{\mu_1}-p_0(\mu_1)s_0)-p_2(\mu_1)(S_{\mu_2}-p_0(\mu_2)s_0)}{p_2(\mu_2)p_1(\mu_1)-p_2(\mu_1)p_1(\mu_2)} \label{yield}
 \end{align}
 where $p_k(\mu)$ is defined by Eqs.(\ref{ps0},\ref{p1},\ref{p2}) and $s_0$, $S_{\mu_1}$, $S_{\mu_2}$ are experimentally observed effective-event rate of $X_{\mu_0}$-windows, $X_{\mu_1}$-windows, $X_{\mu_2}$-windows, and $\mu_0=0$. We also have the following formula for the upper bound value of phase-flip error rate
  of effective ancillary photons of $Z$-windows
  \begin{equation}e_1^{ph}\le{\bar e}_1^{ph}=\frac{S_{\mu_1} E_{\mu_1}^X-e^{-2\mu_1}s_0/2}{2\mu_1 e^{-2\mu_1}s_1}.\label{phd}
 \end{equation}
  If we use infinite intensities, we can even verify the exact value of $s_1$, as was applied in our numerical simulation and other works on TF-QKD.
 %\\ \underline{Note  3} {\em They can just announce all values of $\delta_A,\delta_B$ publicly and the secret channel is not necessary}
\\
\underline{Note 3} {\em quasi-purification}  \\Since {\em their} goal is to have the final key only, a true purification to ancillary photons is not necessary\cite{simple}.
{\em They} can choose to    measure all ancillary photons of $Z$-windows in advance\cite{simple} in photon-number basis and then take virtual purification to classical data of $Z$-windows corresponding to those effective events.  {\em They} then take a virtual quasi-purification to the classical data, which is just the final key distillation. Also, the pre-shared extended state for a $Z$-window is just
 $(|01\rangle\langle 01| \otimes |01\rangle\langle 01| + |10\rangle\langle 10| \otimes |10\rangle\langle 10|)/2$. The pre-arranged restriction of local phase-shifts by Eq.(\ref{cons})  is now trivial and ignored in  $Z$-windows.
\subsection{Protocol 4 and complete SNS protocol}
 \noindent Protocol 4 is exactly equivalent to the simplified SNS protocol, {\em they} need an extended state $\Omega$ as Eq.(\ref{zt0}) for a $Z$-window and pre-share
 a classical information for $Z$-windows and $X$-windows.
 \\4-1 {\em They} send out to Charlie the real photons of state $\Omega$ in a $Z$-window and two-mode coherent state $\rho_X$ as defined in Eq.(\ref{xx}) in an $X$-window.
 \\4-2 {\em They} take post selection for  $X$-windows by the criterion of Eq.(\ref{slice}).
 \\4-3  {\em They} verify the phase-flip error rate $e_1^{ph}$ by the decoy-state analysis. Also, {\em they} verify $n_1$, the number of un-tagged bits
 in $Z$-basis by decoy-state analysis.
 \\4-4  {\em They} each observe the ancillary state for bit value of an effective event in a $Z$-window. {\em They} take error test for $Z$-basis encoding by classical communication.
 \\4-5 After virtual purification to the classical data (final key distillation), {\em they} obtain the final key with the length given by Eq.(\ref{kr1}).
\\ \underline{Note 1} In this protocol, the state $\Omega$ of Eq.(\ref{zt0}) for $Z$-basis is a classical mixture of state $\Omega_1$ of Eq.(\ref{zt1}) and other states.
Given the notes under Virtual protocol 3,  if {\em they} have only used state $\Omega_1$ for $Z$-windows in Virtual protocol 4,  it is equivalent to Virtual protocol 3 which has been shown to be secure already.  We can now apply the tagged model\cite{gllp}. Consider $Z$-windows. Some of the $Z$-windows use the extended states of $\Omega_1$, we name these $Z$-windows as $Z_1$-windows.
%We denote set $c_1$ for all those effective bits due in $Z_1$-windows.
Suppose there are $n_1$ bits from $Z_1$-windows. These $n_1$ bits from $Z_1$-bits are regarded as the {\em un-tagged bits}.
All the other bits corresponding  are regarded as {\em tagged bits}. Applying  the tagged model, {\em they} can distill a secure final key from all bits with   length
\begin{equation}\label{kr1}
n_F = n_1 - n_1H(e_1^{ph}) - n_tH(E_Z)
\end{equation}
where $n_t$ is the number of total raw bits corresponding to effective events and $E_Z$ is the bit error rate in $Z$-basis. An error  bit in $Z$-basis is defined as
the case that Alice's bit value is different from Bob's bit value in an effective $Z$-window. In the formula above, values of $n_1$, $e_1^{ph}$ can be computed by the decoy-state method, while $n_t,\;E_Z$ are directly observed by test.
\\
\underline{Note 2} Equivalence to the real SNS protocol. Suppose in the Virtual protocol 4 above, the pre-shared classical information takes  probability $p_Z$ for a $Z$-window, probability $p_{\mu_k}$ for an $X$-window using intensity $\mu_k$, and $p_Z+\sum_kp_{\mu_k}=1$. In our real protocol, {\em they} each take probability $q_z$ for a signal window and $q_{\mu_k}$
for a decoy window with intensity $\mu_k$. In this way, the real protocol has  a probability $q_Z^2$ for a $Z$-window, $q_{\mu_k}^2$ for an $X_{\mu_k}$-window. Discarding events of all those mismatching windows, the real protocol is equivalent to Virtual protocol 4 above with setting of
\begin{align}
& p_Z =  q_Z^2/\mathcal N,\; p_{\mu_k} =  q_{\mu_k}^2/\mathcal N\\
& \mathcal N = q_Z^2 + \sum_k q_{\mu_k}^2.
\end{align}
But the security of Virtual protocol 4 has already been proven.
On the other hand, we can also construct another Virtual protocol including events of mismatching windows in the real protocol. Suppose in the real protocol, the real-photon state sent-out for a mismatching window is $\rho_{\mathcal M}$.
\\Virtual protocol 5\\
{\em They} pre-share classical information on time windows of $Z$, $\{X_{\mu_k}\}$, and mismatching windows $\mathcal M$, assigning probabilities of $p_Z$, $\{p_{\mu_k}\}$, and $p_{\mathcal M}$ for each of them. {\em They} also pre-arrange the different window commitment  of Alice and Bob  for the mismatching windows, i.e. make sure they have committed differently for all pre-agreed mismatching windows.
\\5-1 {\em They} send out the real photons of state $\Omega$ of Eq.(\ref{zt0}) in a $Z$-window, the two-mode coherent state $\rho_X$ as defined in Eq.(\ref{xx}) in an $X$-window, and state $\rho_{\mathcal M}$ in an mismatching window.
 \\5-2 {\em They} each announce the specific type of window committed and discard those mismatching windows. {\em They} take post selection for $X$-windows by   Eq.(\ref{slice}).
 \\5-3, 5-4 are identical to virtual protocol 4.
 \\ \underline{ Note 1}. The first half of 5-2 is not necessary in protocol 5 itself, but we arrange it in order to show that the real protocol is strictly equivalent to Virtual protocol 5. Explicitly, if we set
 \begin{equation}
 p_Z =  q_Z^2, \; p_{\mu_k} =  q_{\mu_k}^2,\; p_{\mathcal M} = 1- p_Z -\sum_k q_{\mu_k}^2,
\end{equation}
in protocol 5, the real protocol is strictly equivalent to it. This completes the security of SNS protocol.
  \section{Concluding remark}
 In conclusion, following the novel idea of TF-QKD\cite{nature},
  we proposed the sending or not-sending  TF-QKD protocol. Our protocol does not need to announce the phase information of signal pulses and hence the traditional decoy-state formulas can be directly applied.  The single-photon interference is not needed in $Z$ basis thus the error rate in $Z$ basis can be negligibly small. This makes  the protocol be tolerable to a fairly large error rate in $X$ basis where single photon interference must be done. Numerical simulation shows that the protocol can exceed a secure distance of 800 km without misalignment error, and more than 700 km with a misalignment error of $15\%$. Even though the misalignment error for the single-photon interference is as large as $25\%$, the protocol can still reach a secure distance of more than 600 km. Thanks to the revolutionary progress made by TF-QKD proposed in \cite{nature}.
  \section*{Appendix: Eavesdropping scheme based on afterwards announced phase information of signal states.}
 Earlier, we showed that our protocol can  apply the traditional decoy-state method directly because the phase information of signal states is never announced. But, if it were announced and it took a role in bit value, then there were Eavesdropping schemes effectively attacking the secret bits. Here we show this by a specific scheme. Consider the original TF-QKD protocol\cite{nature} as shown in Fig.3. Suppose  coherent state of intensity $\mu$ is used by each sides for signal pulses.  The pulse pairs are phase modulated before being sent out for Charlie. The phase modulation includes the coding phase (0 or $\pi$) at each sides  and the random phase shift we assume to be $\rho$ at both sides\cite{nature}.
  \begin{figure}%[htb]
    \includegraphics[width=200pt]{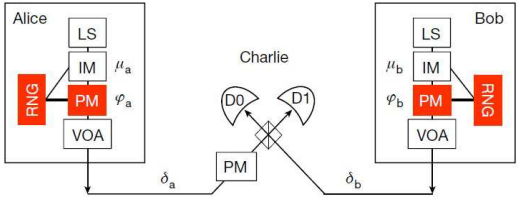}
    \caption{Schematic picture of TF-QKD  taken from \cite{nature}.}
\end{figure} After modulation, the  states of signal pulse pairs are   two-mode coherent states $|\psi^+\rangle=|\sqrt\mu e^{i\rho}\rangle|-\sqrt\mu e^{i\rho}\rangle$ for bit value 0 and $|\psi^-\rangle=|-\sqrt\mu e^{i\rho}\rangle|\sqrt\mu e^{i\rho}\rangle$ for bit value 1, which will cause clicking of detector $D0$ only; and also $|\phi^+\rangle=|\sqrt\mu e^{i\rho}\rangle|\sqrt\mu e^{i\rho}\rangle$ for bit value 0 and $|\phi^-\rangle=|-\sqrt\mu e^{i\rho}\rangle|-\sqrt\mu e^{i\rho}\rangle$ for bit value 1, which will cause the clicking of detector $D1$ only. Note that the strong reference light is controlled by Eve, here we have assumed the reference phase to be 0 for conciseness.  Eve applies the following scheme:
{\bf Step 0}.
 Eve can set whatever channel transmittance. For simplicity, we assume Eve sets the channel transmittance to be 1 here. Consider Fig.1. Before the twin pulses enter the beam splitter, Eve.(Charlie) just honestly does whatever as requested by the the TF-QKD protocol.
  {\bf Step 1} Eve. takes non-destructive crude measurement to project the output light from the beam splitter to vacuum or non vacuum subspace.   Suppose she obtains non-vacuum, she stores the detected state and continue the attacking scheme.
{\bf Step 2}  Eve takes a crude measurement to project the stored state  either to the subspace $\mathcal S=\{|1\rangle, |2\rangle\}$ or to the subspace $\tilde{\mathcal S}=\{|3\rangle,|4\rangle,|5\rangle, \cdots \}$. Suppose the outcome is $\mathcal S$, she stores the state and continues.
{\bf Step 3}  Eve. takes the following unitary transformation to her stored state above:
$
    \ket1 \rightarrow \sqrt {\mu} \ket1 +\sqrt{1-\mu}\ket{m_0},\;
    \ket2 \rightarrow \ket2
    $
where    $\ket{m_0}$ is a state orthogonal to both $\ket1$ and $\ket2$. Eve. takes a crude measurement which collapses the stored state in Step 3 either to state $|m_0\rangle $ or the subspace $\mathcal S$ spanned by the Fock states $\{|1\rangle,|2\rangle\}$. Suppose she obtains subspace $\mathcal S$ in step 3, she stores the state and announces which detector ($D0$ or $D1$) has counted.  She wait until  Alice and Bob's announcement, then goto Step 5.

Note: until now we always assume Eve obtains the results in favor of her attacking in those non-trace preserving maps. The point is that, at any step, if Eve doesn't obtain the measurement outcome in favor of her,  she just announces  that she has not detected anything.

{\bf Step 5}  After Alice and Bob announce the value of $\rho$, bases of each pulse pairs, and which pulses are decoy pulses and which pulses are signal pulses,  Eve. can takes a phase shift operation  to  her stored state, changing it into one of the following 2 states corresponding on  bit value 0 or 1 of the incident pulse pair:
$
\frac{1}{\sqrt 2}(|1\rangle\pm |2\rangle).
$
This enables Eve. to know the bit value for sure without causing any noise  by a projective measurement.

 Here are details of the state evolution for the non-trace-preserving map above. Suppose at Step 1 detector $D0$ counts only, the incident state can be either $|\psi^+\rangle$ or  $|\psi^-\rangle$. If the incident state is $|\psi^+\rangle$, the stored states $\{|\psi^+_i\rangle\}$ at the end of each Steps $\{i\}$ are:
 $
|\psi_1^+\rangle
=\mathcal N_1 \sum_{k=1}^\infty \frac{(\sqrt {2\mu}e^{i\rho})^k}{\sqrt{k!}}|k\rangle;
$,
$
|\psi_2^+\rangle = \mathcal N_2 (\sqrt{\mu}|1\rangle+\mu e^{i\rho}|2\rangle);
$,
$
|\psi_4^+\rangle = \frac{1}{\sqrt 2} (|1\rangle+e^{i\rho}|2\rangle).
$
$
 |\psi_5^+\rangle = \frac{1}{\sqrt 2}(|1\rangle + |2\rangle).
 $
All parameters $\mathcal N_1,\mathcal N_2,\mathcal N_4$ are normalization factors.

 Similarly, given the incident states $\{|\psi^-\rangle\}$, we can also calculate time evolution of $\{|\psi^-\}\rangle$  at each Steps $\{i\}$, and we obtain:
$
 |\psi_5^-\rangle = \frac{1}{\sqrt 2}(-|1\rangle + |2\rangle).
 $
 This means $|\psi^+_5\rangle$ and $|\psi^-_5\rangle$ are orthogonal to each either and Eve can know the corresponding bit value for sure.
In the same way, one can easily show that Eve can also obtain full information of bit values without causing disturbance.

  In the Eavesdropping above, the fraction of bits caused by single-photon state is $50\%$ among all raw bits.  According to the key rate formula (Eq.(2)) of Ref.\cite{nature}, TF-QKD will present a key rate of $50\%$ from raw key to final key although the actual key rate is obviously 0. This means  the key rate formula does not match the protocol itself there. The root of the problem is that Eve can make use of post announced phase information of signal states there. Given that protocol, one have to apply a different key rate formula.
  \\{\em Note added}: After we announced our Eavesdropping scheme  on the arXiv:1805.02272, it was then suggested  using different key rate formulas directly pointing to non-random-phase coherent states \cite{xf,tmk}.


\begin{thebibliography}{99}
\bibitem{note1} The data of time windows other than $Z$-windows and $X$-windows, e.g., a time window when one party decides to send a decoy pulse and the other party decides not-sending can offer more data for the decoy-state analysis.
This is useful in non-asymptotic strudy of the decoy-state method applycation.
\end{thebibliography}
\end{document}